\documentclass[preprint,amsmath,amssymb,superscriptaddress,notitlepage]{revtex4-1}
\usepackage{epsfig}
\usepackage{graphicx}
\usepackage{color}
\usepackage{bm}
\usepackage{mhchem}
\usepackage{miller}

\def\myhighlight#1{\textcolor{black}{#1}}

\usepackage[utf8]{inputenc}  

\begin{document}

\title{Anisotropic magneto-thermal transport in Co$_2$MnGa thin films}

\author{Philipp Ritzinger}
\affiliation{Institut f{\"u}r Festk{\"o}rper- und Materialphysik and W{\"u}rzburg-Dresden Cluster of Excellence ct.qmat, Technische Universit{\"a}t Dresden, 01062 Dresden, Germany}
\email[Corresponding author: ]{philipp.ritzinger@web.de}
\author{Helena~Reichlova}
\affiliation{Institut f{\"u}r Festk{\"o}rper- und Materialphysik and W{\"u}rzburg-Dresden Cluster of Excellence ct.qmat, Technische Universit{\"a}t Dresden, 01062 Dresden, Germany}
\author{Dominik~Kriegner}
\affiliation{Institute of Physics ASCR, v.v.i., Cukrovarnick\'a 10, 162 53, Praha 6, Czech Republic}
\affiliation{Max Planck Institute for Chemical Physics of Solids, N\"othnitzer Stra\ss e 40, 01187 Dresden, Germany}
\author{Anastasios~Markou}
\affiliation{Max Planck Institute for Chemical Physics of Solids, N\"othnitzer Stra\ss e 40, 01187 Dresden, Germany}
\author{Richard~Schlitz}
\affiliation{Leibniz Institute for Solid State and Materials Research Dresden (IFW Dresden), Institute for Metallic Materials, 01069 Dresden, Germany}
\affiliation{Institut f{\"u}r Festk{\"o}rper- und Materialphysik and W{\"u}rzburg-Dresden Cluster of Excellence ct.qmat, Technische Universit{\"a}t Dresden, 01062 Dresden, Germany}
\author{Michaela~Lammel}
\affiliation{Leibniz Institute for Solid State and Materials Research Dresden (IFW Dresden), Institute for Metallic Materials, 01069 Dresden, Germany}
\author{Gyu Hyeon Park}
\affiliation{Leibniz Institute for Solid State and Materials Research Dresden (IFW Dresden), Institute for Metallic Materials, 01069 Dresden, Germany}
\author{Andy Thomas}
\affiliation{Leibniz Institute for Solid State and Materials Research Dresden (IFW Dresden), Institute for Metallic Materials, 01069 Dresden, Germany}
\author{Pavel~St\v reda}
\affiliation{Institute of Physics ASCR, v.v.i., Cukrovarnick\'a 10, 162 53, Praha 6, Czech Republic}

\author{Claudia~Felser}
\affiliation{Max Planck Institute for Chemical Physics of Solids, N\"othnitzer Stra\ss e 40, 01187 Dresden, Germany}
\author{Sebastian~T.~B.~Goennenwein}
\affiliation{Institut f{\"u}r Festk{\"o}rper- und Materialphysik and W{\"u}rzburg-Dresden Cluster of Excellence ct.qmat, Technische Universit{\"a}t Dresden, 01062 Dresden, Germany}
\affiliation{Department of Physics, University of Konstanz,
  Universitatsstrasse 10, 78457 Konstanz, Germany}
\author{Karel~V\'yborn\'y}
\affiliation{Institute of Physics ASCR, v.v.i., Cukrovarnick\'a 10, 162 53, Praha 6, Czech Republic}

\date{Dec22, 2020}

\begin{abstract}
Ferromagnetic \ce{Co2MnGa} 
has recently attracted
significant attention due to effects related to non-trivial topology of its
band structure, however a systematic study of canonical
magneto-galvanic transport effects is missing. Focusing on high quality 
thin films, here we systematically
measure anisotropic magnetoresistance (AMR) and its thermoelectric
counterpart (AMTP). We model the AMR data by free energy
minimisation within the Stoner--Wohlfarth~formalism and conclude that
both crystalline and non-crystalline components of this
magneto-transport phenomenon are present in \ce{Co2MnGa}. Unlike the AMR
which is small in relative terms ($\sim 0.1\%$), the AMTP is large due
to a change of sign of the Seebeck coefficient as a function of
temperature. This fact is discussed in the context of the Mott rule
and further analysis of AMTP components is presented.


\end{abstract}


\maketitle
\pagebreak

Electron transport phenomena in magnetically ordered materials span a
vast range both historically and from the point of view of
complexity. While some of them which have been known for a long time,
such as the anisotropic magnetoresistance (AMR)~\cite{Thomson:1857_a},
remain a subject of roughly constant interest until
today~\cite{Wang:2020_a,Volny:2020_a,Miao:2020_a} 
others rose to prominence only recently. Such is the case of the
anomalous Nernst effect~\cite{Wesenberg:2018_a} (ANE) for example, an
outstanding member of the field of spin caloritronics~\cite{Yu:2017_a}.
\myhighlight{
In a typical thin film geometry with magnetic field applied in the
direction normal to the film plane and a thermal gradient in
the sample plane, the ANE signal is detected in the other (perpendicular)
in-plane direction.} This effect is particularly strong in ferromagnetic
\ce{Co2MnGa}~\cite{Hu:2020_a} (identified as a Weyl semimetal~\cite{Sakai2018})
and having thus drawn considerable interest it has been investigated
in sufficient detail already~\cite{Reichlova2018, Belopolski2019, Park2020}.

In this work, we extend the discussion also to effects which occur when magnetic field is applied in the sample plane. The studied thin film of Heusler alloy \ce{Co2MnGa} represents an ideal model system because of its high crystalline quality, relatively strong magneto-thermal response, its high Curie temperature ($T_C=694$~K) and high spin polarisation \cite{Graf:2011_a}. We study systematically the magneto-thermal transport response when the magnetic field is rotated in three perpendicular rotation planes.
Along with AMR, the anisotropic magneto-thermopower (AMTP) is
reported and compared qualitatively. We discuss the applicability of a simple
Stoner--Wohlfarth-based model (as used recently in a different context,
for example in Ref.~\cite{Volny:2020_a}) to the AMTP data and compare
the ratio of amplitudes of AMTP and AMR in various samples and at different
temperatures. We report crystalline contributions to both AMR and AMTP and
interestingly, some of these seem missing (or at least they are
significantly weaker) in the latter effect. Inconsistencies related to
the straightforward application of the Mott rule to our measurements
suggest a sizable phonon~\cite{Protik:2020_a} or
magnon~\cite{Polash:2020_a} drag contribution to the thermopower.

This paper is structured as follows: in Sec I we are motivating the
comparison of anisotropic magnetoresistance (AMR) and anisotropic
magnetothermopower (AMTP) and provide some background on these two effects.
In Sec II, the formalism used for the data analysis is
introduced and sample fabrication and characterisation is described in
Sec. III. Finally, experimental results are shown and discussed in Sec. IV;
an outlook and and summary is provided in Sec. V.

\subsection*{I. INTRODUCTION}

Both AMR and AMTP refer to voltage variation as a function of the
angle between magnetization and the driving force (electrical current
$\textbf{j}$ in case of AMR and temperature gradient $\nabla T$ in
case of AMTP) or the angle between magnetization and the crystal axes
in the respective setting. Although they could probe similar physical
properties, there are many more reports about AMR than about AMTP,
mostly due to experimental challenges in measuring AMTP. In the
following, we give a brief introduction of both.

Since the original observation of AMR by William Thomson
\cite{Thomson:1857_a} the AMR is typically understood as a variation
of resistance as a function of magnetisation $\textbf{M}$ direction
\cite{Rushforth:2007_a} $\Delta\rho_{yy}(\varphi) = C_I \cos 2\varphi$,
with $\varphi$ denoting the angle between current and magnetisation. However, there is another level of complexity in AMR. In 1938, the discussion of this effect was extended to the influence of crystalline symmetry.  W. Doering~\cite{Doring:1938_a}
carried out symmetry-based AMR analysis \cite{Ranieri:2008_a} of resistivity tensor using a
series expansion up to fourth order in powers of the direction cosines of the
magnetization. These expansions contain AMR terms different from the
"non-crystalline" $\cos 2\varphi$ ones and such additional terms are
sometimes called "crystalline" AMR since they reflect the crystal
symmetry and not  the symmetry breaking induced by the electrical
current direction. Consequently, unlike the non-crystalline AMR, the
crystalline AMR contributions can be non-zero even if $\varphi=\pi/2$
remains constant (for example, during the magnetic field
rotation in the plane perpendicular to $\textbf{j}$. Such a situation
will be discussed in the experimental setup sketched in
Fig.~\ref{Fig2}(e) below. 

\myhighlight{
AMTP is the thermoelectrical counterpart of AMR. The basic phenomenon
(voltage drop induced by a temperature gradient)
was discovered by T.J. Seebeck already in 1821, thus establishing the field 
of thermoelectrics. Hints at its anisotropy came much later~\cite{Ky:1966_a}
however and since then, AMTP has attracted relatively small attention
compared to the AMR. Nevertheless, increasing global demand for energetically
sustainable solutions~\cite{Snyder:2008} and the need of advanced
microscopy techniques~\cite{Janda:2020_b} sparked new interest in this effect.
}


Recent thermoelectric studies in solid state magnetism focus mostly on
the evaluation of $\Sigma$ (the Seebeck coefficient~\cite{Loevvik:2020_a})
or the anomalous Nernst effect (ANE)~\cite{Yu:2017_a}. The Seebeck coefficient
provides information about the charge carriers, such as concentration,
effective mass or dominant type (electrons or holes) and ANE stirs interest
\myhighlight{due to the connection to band structure topology~\cite{note2}}
and better technological prospects 
in thermoelectric energy harvesting~\cite{Sakai2020}.
In addition to the ANE, the Seebeck coefficient anisotropy comes with the
possibility for its tensor components $\Sigma_{xy}$ and
$\Sigma_{yy}$ to depend on the magnetic field direction: the
anisotropic magneto-thermopower (AMTP) which is a direct analog to the
non-crystalline AMR, can be expressed as
\begin{eqnarray} \nonumber
  \Sigma_{\mathrm{xy}} &=&S_I\sin 2\varphi\qquad \mbox{ accompanied by }\\
  \Delta\Sigma_{\mathrm{yy}}  &=&S_I\cos 2\varphi, \label{eq1}
\end{eqnarray}
\myhighlight{
where $\Delta\Sigma_{\mathrm{yy}}$ is the difference between $\Sigma_{yy}$
and its average. These relations pertinent to polycrystalline
materials can be straightforwardly derived by angular averaging the
Seebeck tensor as given by Eq.~3.31 in Ref.~\cite{Limmer:2008_a} and
we also discuss this relationship in Sec.~III.}

In contrast to ANE, the AMTP is rarely a topic of systematic studies;
with few exceptions~\cite{Althammer:2012,Reimer:2017_a} reports are
usually limited to assume its existence (in
longitudinal~\cite{Jayathilaka:2015_a,Janda:2020_a}
or transversal~\cite{Ky:1966_a,Avery:2012_a}
geometry also known as the planar Nernst effect) and the AMTP often
assumes the role of an unwanted artefact. The main reasons are the
notorious difficulty to precisely quantify direction and amplitude of
a thermal gradient and small magnitude of the measured
thermo-voltages, typically on  resolution limit of a common laboratory
equipment.  This makes the AMTP experiment significantly more
challenging than a simple resistivity measurement with a well-defined
current direction. The thermal gradient quantification is even more
complex in thin film samples where the substrate acts as a heat
sink. The lack of detailed understanding of AMTP becomes
obvious when considering systems with various contributions to AMR.
In particular, very few reports show more complex symmetry of AMTP~\cite{Althammer:2012}, the existence of a crystalline component in the AMTP is not yet established and a comparison between the crystalline contributions to AMR and AMTP is entirely missing. The understanding of AMTP is not only a fundamental scientific question, but it is equally important in order to exclude various artefacts in experiments during which thermal gradient is unintentionally generated.

%


%

\subsection*{II. SAMPLE FABRICATION AND CHARACTERIZATION}

\begin{figure}[h]
	\includegraphics[scale=0.55,angle=-90]{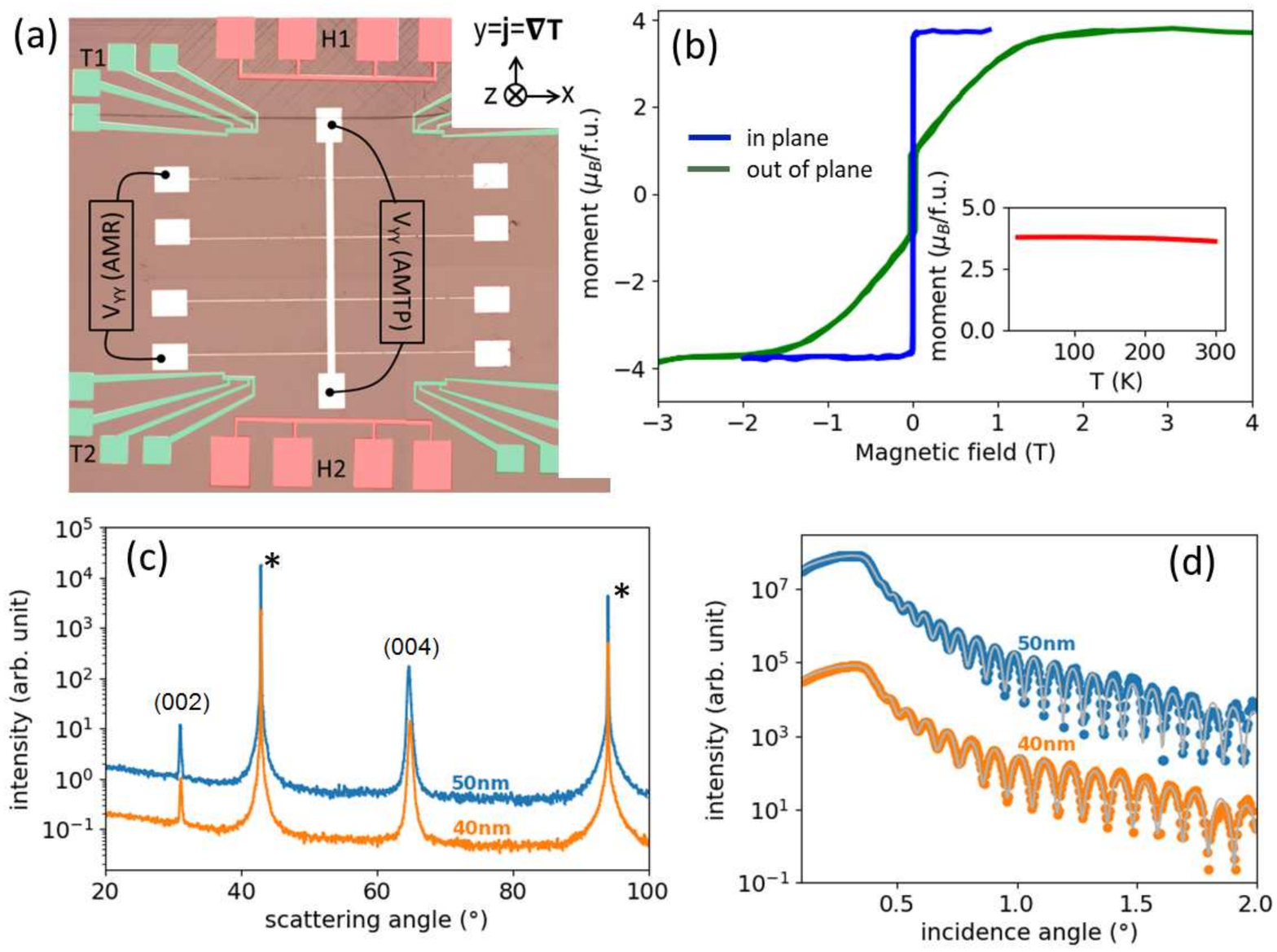}

\caption{Sample characterization. a) Schematic image of the
  sample. The white areas are \ce{Co2MnGa}. $V_\mathrm{xx}$ (AMR) are
  the contacts used for measusing voltage in the AMR experiment, while
  the current $I$ was applied between $V_\mathrm{xx}$ (AMTP). In the
  AMTP experiment, the voltage was measured at $V_\mathrm{xx}$
  (AMTP). The red colored areas are \ce{Pt} wires which are used in
  the AMTP experiments as on-chip heater and thermometer (green). b)
  in-plane and out-of-plane measurement of the magnetization
  vs. applied magnetic field. The inset show the magnetization
  vs. temperature. The saturated magnetic moment is approx. $4
  \mu_{\mathrm{B}}$ per formula unit. c) X-ray diffraction radial scans of Co2MnGa films with 40 and 50nm thickness. Bragg peaks labeled by an asterisk (*) originate from the (00L) series of the MgO substrate, while Bragg peaks of Co2MnGa are labeled by their Miller indices. d) X-ray reflectivity pattern of the 40 nm and 50 nm thick Co2MnGa films. Experimental data are shown as data points, while the solid line represents a model fit based on the extended Parratt formalism.}
	\label{Fig1}
\end{figure}

The Co$_2$MnGa thin-film samples are fabricated by magnetron sputtering on
MgO(001) substrates using a multisource Bestec UHV deposition system
from Co, Mn and MnGa sputter targets. Growth and post-growth annealing
was performed at 500$^\circ$C.
After the Co$_2$MnGa thin-film growth, 3~nm of Al were deposited at room
temperature to prevent oxidation. Further details of the growth
procedure can be found elsewhere~\cite{Markou:2019}. Here, two samples
showing highest crystal quality with Co$_2$MnGa thickness of 40~nm and
50~nm are studied.
The chemical composition and structural investigation conducted by X-ray
diffraction techniques showed Bragg peaks corresponding to the material
composition revealing a high degree of atomic order similar to
Refs.~\cite{Reichlova2018,Markou:2019}.
Figure~\ref{Fig1}(c) shows the symmetric radial X-ray diffraction scans,
which includes diffraction from lattice planes parallel to the substrate
surface. Given the epitaxial alignment of~Co$_2$MnGa(001)[110]
$\parallel$ MgO(001)[100], i.e. an in-plane~45$^\circ$ degree rotation,
only the $00L$ Bragg peaks are visible in Fig.~\ref{Fig1}(c).
Well defined, narrow Bragg peaks evidence the good chemical homogeneity and
crystal quality. While bulk Co$_2$MnGa has a cubic L2$_1$ crystal structure
the thin-films exhibit an
epitaxial strain-induced tetragonal distortion with slight contraction
along the out of plane [001] direction. Resulting $c/a$ ratio is around
0.99~\cite{Markou:2019}.
Figure~\ref{Fig1}(d) shows X-ray reflectivity data of the~40~nm and
50~nm thick Co$_2$MnGa
epilayer displaying Kiessig~fringes that extend beyond the measurement
range. This bears witness to a low surface and interface roughness which were
determined to be below 7~\AA{} by modelling using an extended Parratt
formalism~\cite{Pietsch:2004}.

Magnetization of these epilayers was measured in a SQUID magnetometer,
which is shown in Fig.~\ref{Fig1}(b). The saturation magnetic moment of about
4 $\mu_\mathrm{B}$ / f.u. is consistent with literature (saturation magnetisation $M_{\mathrm{sat}}$ 720 kA/m) 
\cite{Reichlova2018,Park2020}.
The films were patterned into 40$\mu$m wide Hall bars by optical
lithography and by a combination of \ce{HCl} and Ar/O2 plasma
etching. A schematic image of the sample is shown in
Fig. \ref{Fig1}(a). After the etching, the heater and thermometers were
fabricated in a lift-off process with 30 nm of sputtered
\ce{Pt}. Platinum wires, highlighted as pink areas in Fig.~\ref{Fig1}(a),
at the top of the Hall bar serve as on-chip heater, while platinum
wires at the side work as an on-chip thermometer (green areas).


\subsubsection*{Experimental Setup}

In the AMR experiments, a sufficiently strong magnetic field ($\mu_0 H= 1.5$~T
stronger than $M_{\mathrm{sat}}\approx 0.9$ T and the anisotropy field
$\approx 0.6$ T corresponding to $k_u$) is rotated in three rotation
planes called XY, ZY and ZX plane as shown in Fig. \ref{Fig2} (a), (c)
and (e). Current $I = 0.2$ mA (corresponding to current density
$1.2\times 10^6$ A/cm$^2$) is applied along the y-axis. At each step
of the magnetic field rotation, data is collected for $I = +0.2$ mA
and for $I = -0.2$ mA and averaged, in order to cancel out the
thermoelectric contributions, which occur in form of voltage
offsets. The experiments are conducted at different temperatures
between 10 K and 300 K. The data are shown as symbols in Fig.~\ref{Fig2}.

In the AMTP experiments, magnetic field is rotated in the same rotation planes XY, ZY and ZX. The experiments are conducted at several temperatures between 100 and 300~K; at lower temperatures the AMTP signal decreases below the resolution owing to the decrease of entropy.

The thermometers are first calibrated by sweeping the temperature of the cryostat from low temperature to room temperature and using a Cernox thermometer in the cryostat as a reference. At each studied temperature, a constant current of $I_\mathrm{thermom} = 0.1$ mA is applied on the on-chip thermometers (platinum wires, green areas in Fig 1), 
while the measured voltage serves as measure of the temperature. Thermal gradient is generated by Joule heating of the heater, typically we apply a current of $I_\mathrm{heater} = 17.5$ mA. The on-chip thermometry allowed us to determine the thermal gradient $\nabla T$, 
which is 0.4 K/mm at $T = 100$ K and 0.5~K/mm at 200 K - 300
K~\cite{Reichlova2018}. In order to reduce noise, the magnetic field
was rotated several times and at each rotation step, several voltage
measurements are taken. In addition, the presented data are averaged
over several magnetic field rotations.  Since the thermal gradient takes long time to stabilise, it was not reversed at each step of the  rotation as in the case of the AMR.

\subsection*{III. PHENOMENOLOGICAL MODEL}

The phenomenological model used in this work was previously employed by
Limmer~et~al.~\cite{Limmer:2006_a,Limmer:2008_a} for AMR in
\ce{(Ga,Mn)As} and extended to AMTP in the same material system by
Althammer~\cite{Althammer:2012} and we present a brief summary here.
Similar schemes are used also in the context of AMR in
antiferromagnets.~\cite{Volny:2020_a}

Coordinate system is chosen as follows: $\textbf{z}$ is the surface
normal vector, which is in the \hbox{[0 0 1]} direction, electric field and
thermal gradient are applied along \hkl[1 1 0] denoted by $\textbf{y}$
and $\textbf{x} = \textbf{y} \times \textbf{z}$. A sketch of the
Hall bar with the coordinate system is shown in Fig \ref{Fig1}(a).

The basic simplifying assumption is that magnetisation $\textbf{M}$ is
saturated (which is plausible given the very narrow hysteresis loop
shown in Fig.~\ref{Fig1}(b) and that we are in a single-domain
state. Stoner-Wohlfarth (SW) model~\cite{SW48} can then be used to infer
the magnetisation direction $\textbf{m} = \textbf{M}/M_\mathrm{sat} $
(here, $M_\mathrm{sat}$ is the saturation magnetisation) for any given
applied magnetic field $\textbf{H}$ by minimizing the free energy density $F$.
We note that this approach is capable of reproducing hysteresis
effects but these never occur in the parameter range of interest
here. Both the Zeeman energy and magnetic anisotropies contribute to $F$
and the latter in \ce{Co2MnGa} shows a cubic anisotropy $k_\mathrm{c}$
and an uniaxial anisotropy $k_\mathrm{u}$ which is expected due to
demagnetization energy and substrate-induced strain. With $\mu_0$
being the vacuum permeability, we use $F = -\mu_{0} \textbf{H}
\boldsymbol{\cdot} \textbf{M} + k_{\mathrm{u}} m_\mathrm{z}^2$ because  
the cubic anisotropy can be neglected (FMR
measurements~\cite{Swekis:2020_a} show that it is two orders of
magnitude smaller than the applied magnetic field).



The resistivity tensor is obtained by making a series expansion in
powers of 
cartesian components of $\textbf{m}$ up to the fourth order. This ansatz was
first developed by Birss and Muduli et al. and applied for
example~\cite{Limmer:2006_a} to \ce{(Ga, Mn)As}. The tensor writes as:
\begin{equation}
	\rho_{\mathrm{ij}}(\textbf{m}) = \rho_{\mathrm{ij}}^{(0)} +  \rho_{\mathrm{ijk}}^{(1)} m_{\mathrm{k}} +  \rho_\mathrm{{ijkl}}^{(2)} m_{\mathrm{k}} m_\mathrm{{l}} +  \rho_{\mathrm{ijklm}}^{(3)} m_{\mathrm{k}} m_{\mathrm{l}} m_{\mathrm{m}} + \rho_{\mathrm{ijklmn}}^{(4)} m_{\mathrm{k}} m_{\mathrm{l}} m_{\mathrm{m}} m_{\mathrm{n}} + ...
\label{Tensor}
\end{equation}
where $\rho_{\mathrm{ij}}^{(0)}$, $\rho_{\mathrm{ijk}}^{(1)}$,
$\rho_{\mathrm{ijkl}}^{(2)}$, $\rho_{\mathrm{ijklm}}^{(3)}$ and
$\rho_{\mathrm{ijklmn}}^{(4)}$ are the expansion coefficients and
magnetisation direction components
$m_{\mathrm{n}} \in \{m_{\hkl[1 0 0]}, m_{\hkl[0 1 0]}, m_{\hkl[0 0 1]} \}$.

The number of independent parameters is reduced owing to
$m_{\mathrm{k}} m_{\mathrm{l}} = m_{\mathrm{l}} m_{\mathrm{k}}$,
the Onsager relation $\rho_{\mathrm{ij}}(\textbf{m}) =
\rho_{\mathrm{ji}}(-\textbf{m})$ and Neumann's principle~\cite{note3}
pertaining to the crystal symmetry. The last mentioned is tetragonal
in our case, whereas the tetragonal
axis is in $z$-direction, since the thin-film samples are strained by
the \ce{MgO} substrate (see Sec. II). The complete form of the
resistivity tensor is not needed for the further process and can be
found in the Appendix. The longitudinal resistivity
$\rho_\mathrm{long}$ is obtained by projecting the resistivity tensor
$\rho$ along the current direction by making use of Ohm's law
$\textbf{E} = \rho \cdot \textbf{J}$ and $E_\mathrm{long} = \textbf{j}
\cdot \textbf{E}$, where $\textbf{J}$ is the current density vector
and $\textbf{j} = \textbf{J}/J$ is the corresponding unit vector, in
our case $\textbf{j} = \frac{1}{\sqrt{(2)}} (1, 1, 0) =
\textbf{y}$. The projection writes as:
\begin{equation}
\rho_\mathrm{long} = \rho_\mathrm{yy} =  \textbf{j} \cdot \rho \cdot \textbf{j}
\label{Proj}
\end{equation}
The longitudinal resistivity $\rho_{\mathrm{yy}}$ in our configuration
is therefore given by:
\begin{equation}
\rho_{\mathrm{yy}}  = \rho_\mathrm{{0}} + a_\mathrm{y2} \cdot m_\mathrm{y}^2
+ a_{\mathrm{z2}} \cdot m_\mathrm{z}^2
+ a_{\mathrm{y4}} \cdot  m_\mathrm{y}^4
+ a_\mathrm{{z4}} \cdot m_\mathrm{z}^4
+ a_\mathrm{{zy2}} \cdot  m_\mathrm{z}^2 \cdot  m_\mathrm{y}^2
\label{AMR}
\end{equation}
where $\rho_0$ is the offset resistivity, $a_\mathrm{y2}$ and $a_\mathrm{z2}$ are the coefficients of the lowest-order AMR terms, $a_\mathrm{y4}, a_\mathrm{zy2}$ and $a_\mathrm{z4}$ are the coefficients of the higher-order AMR terms and $m_\mathrm{y}$ and $m_\mathrm{z}$ are the y- and z-component of $\textbf{m}$ in the coordinate system $\{x, y, z \}$ as introduced above.

The derivation of the longitudinal Seebeck coefficient
$\Sigma_\mathrm{yy}$ is analogous to the resistivity. The only
difference is here that the Onsager relation connects the Seebeck
tensor with the Peltier tensor and thus cannot be used to reduce the
number of independent parameters. Hence, $\Sigma_\mathrm{yy}$ contains
an additional term:
\begin{equation}
\Sigma_{\mathrm{yy}} = \Sigma_{0} + s_{\mathrm{y2}} \cdot  m_\mathrm{y}^2
+ s_{\mathrm{z2}} \cdot  m_\mathrm{z}^2
+ s_{\mathrm{zyx}} \cdot m_\mathrm{z} \cdot m_\mathrm{y} \cdot  m_\mathrm{x}
+ s_{\mathrm{zy2}} \cdot  m_\mathrm{z}^2 \cdot  m_\mathrm{y}^2
+ s_{\mathrm{y4}} \cdot m_\mathrm{y}^4
+ s_{\mathrm{z4}} \cdot m_\mathrm{z}^4
\label{AMTP}
\end{equation}
where analogously $\Sigma_0$ is the thermoelectric offset,
\myhighlight{$s_\mathrm{y2}$ and $s_\mathrm{z2}$ are the coefficients} of the
lowest-order AMTP terms and $s_\mathrm{y4}$, $s_\mathrm{z4}$,
$s_\mathrm{zy2}$ and $s_\mathrm{zyx}$ are the coefficients of the
higher-order AMTP terms. Since the magnetic field is going to be
rotated in either the XY-, the ZY- or the ZX-plane and amongst the anisotropies only the out-of-plane uniaxial term is significant, one of the 
$m_\mathrm{i}$ is in every plane expected to be zero
(e.g. $m_\mathrm{x}$ in the ZY-plane), the term $m_\mathrm{z} \cdot
m_\mathrm{y} \cdot  m_\mathrm{x}$ is expected to be zero in every of
our rotation planes and is thus ignored. Hence, the AMR and AMTP
formulae contain the same terms {\em in our measurement setup.} \myhighlight{
We note that for polycrystals, only the first two terms in Eq.~(\ref{AMTP})
remain and moreover, $\frac12 s_{\mathrm{y2}}=S_I$ Eq.~(\ref{eq1}). In
other words, all other terms in Eq.~(\ref{AMTP}) can be classified as
crystalline AMTP.}

In the following, we will analyse experimental AMR and AMTP data using
Eqs.~(\ref{AMR}) and~(\ref{AMTP}) combined with the SW model which provides
a link between external magnetic field 
and magnetisation $\mathbf{m}=(m_\mathrm{x},m_\mathrm{y},m_\mathrm{z})$
that enters those equations. The final results of the fitting
procedure are depicted in Fig.~\ref{Fig2}. It will be shown in
Fig.~\ref{Fig2res} that for AMR, all terms in Eq.~(\ref{AMR}) need to be
retained lest the quality of fits deteriorate significantly in some measurement
configurations. On the other hand, 
the last three terms of Eq.~(\ref{AMTP}) are not needed for a good fit
of AMTP; $s_\mathrm{zyx}$ cannot be inferred from our data as already mentioned.

 \subsection*{IV. RESULTS}

Experimental data (symbols) and fits using the phenomenological model
(lines) for both AMR and AMTP in the 50~nm sample are shown in
panels (b), (d) and (f) of Fig.~\ref{Fig2}. While the AMR and AMTP data,
with suitable scaling, seem alike in panels (b) and (d), the rotation of
$\textbf{M}$ in the plane perpendicular to $\textbf{j}$ (see Fig.~\ref{Fig2}f)
gives a different picture. We elaborate on this finding below and only
note here, that in the latter configuration, non-crystalline
terms~\cite{Rushforth:2007_a} do not contribute to the measured AMR
and AMTP which will now be discussed separately.

\subsubsection*{AMR}

\begin{figure}[h]
\includegraphics[scale=0.65,angle=-90]{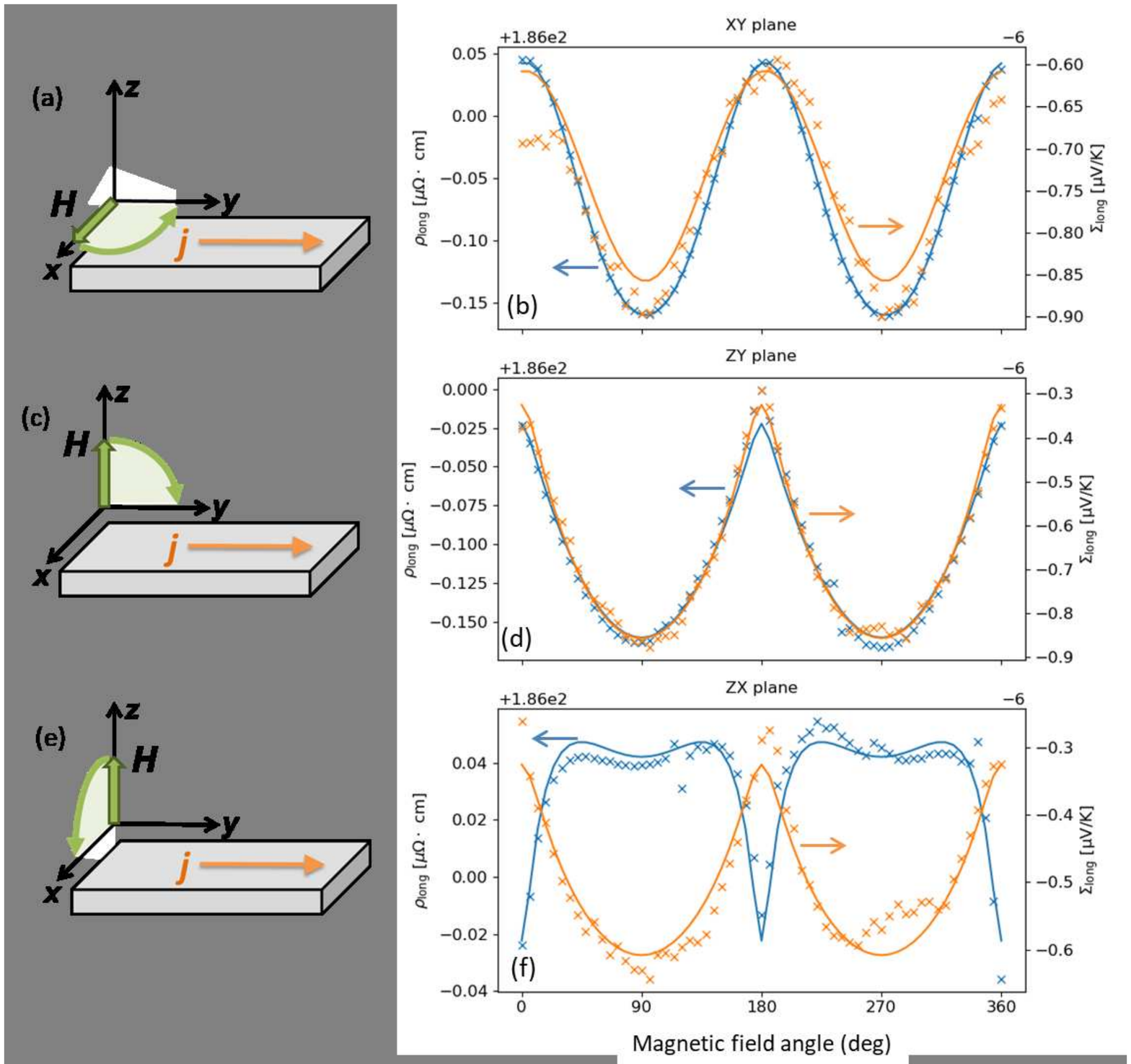}
\caption{Results of the AMR and AMTP measurement (crosses) and fitting
  (continuous lines) in the 50
  nm sample at 300 K (AMR) and 250 K (AMTP). a) b), c) Schematic
  sketch of the magnetic field rotation in the XY, ZY, ZX plane with
  respect to the coordinate system and the sample. d), e), f) AMR
  (blue, left y-axes) and AMTP (orange, right y-axes) results; note
  the shifts of vertical axes indicated at the top of each panel:
  minimum resistivity, which occurs for $\textbf{m}\parallel\textbf{j}$,
  is 171~$\mu\Omega$.cm.}
\label{Fig2}
\end{figure}

\begin{figure}[h]
        \includegraphics[scale=0.56]{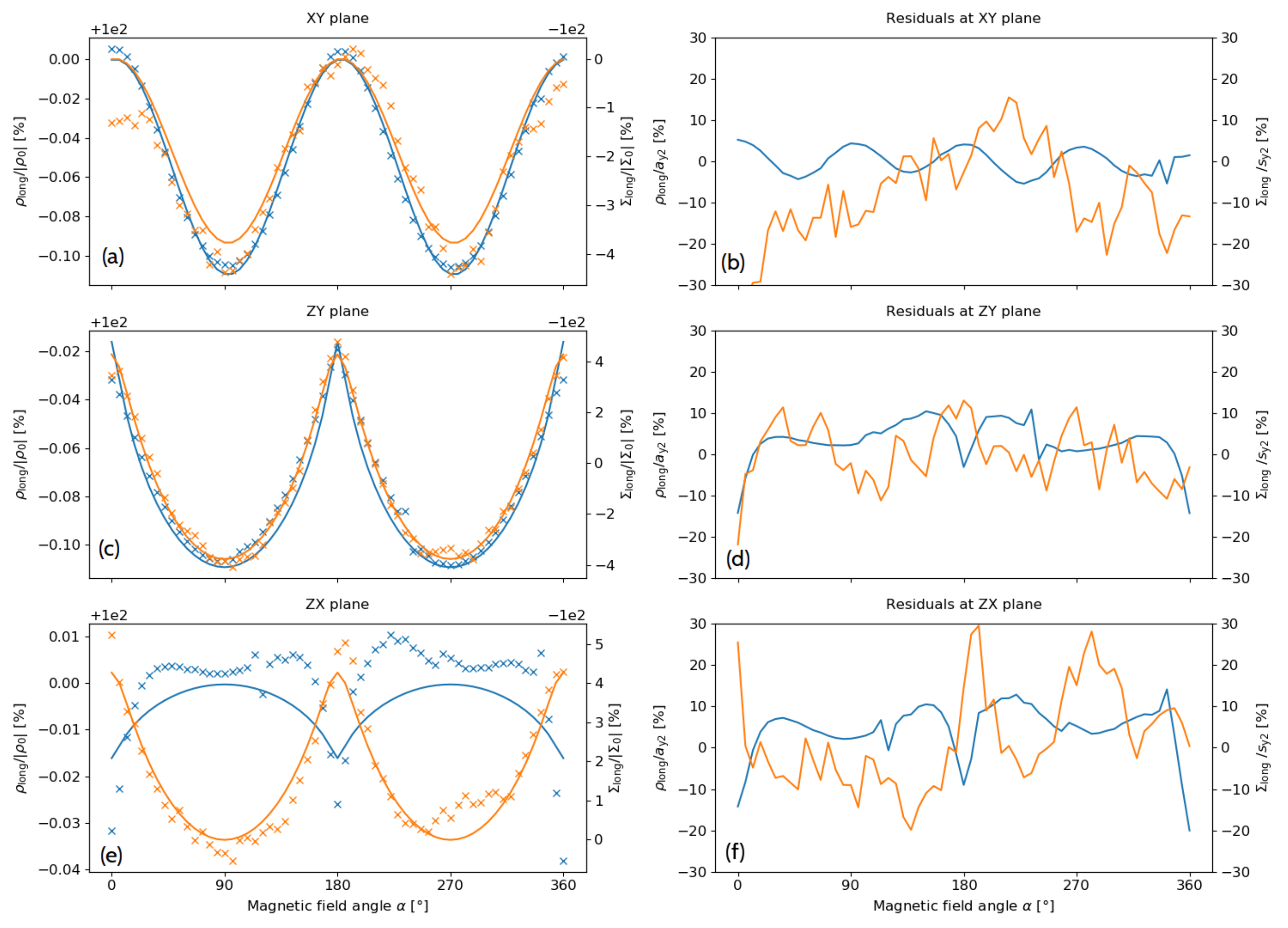}
	\caption{ (a,c,e) Data from Fig.~\ref{Fig2} where AMR
          (AMTP) is fitted by Eq.~\ref{AMR} (Eq.~\ref{AMTP}, respectively)
          with second-order terms only; a lower quality, compared to
          Fig.~\ref{Fig2}(f), of the AMR fit in the bottom panel is clearly
          apparent but systematic deviations can also be observed for the XY
          and ZY rotations.  (b,d,f) The difference of the
          experimental data and fits on the left (fit
          residuals), see text for a detailed discussion.
        }
	\label{Fig2res}
\end{figure}

The phenomenological fit to AMR data (blue crosses in Fig.~\ref{Fig2}) takes
into account the uniaxial magnetic anisotropy $k_\mathrm{u}$,
lowest-order terms $a_\mathrm{y2}$ and $a_\mathrm{z2}$ and also
higher-order AMR terms $a_\mathrm{y4}$, $a_\mathrm{z4}$ and
$a_\mathrm{zy2}$. Eq.~(\ref{AMR}) combined with $k_u$ of the SW model resulted
in a very good agreement between the data and model. On the other hand,
fits omitting the higher-order crystalline terms (specifically,
$a_\mathrm{y4}$, $a_\mathrm{z4}$ and $a_\mathrm{zy2}$) shown on the
left of Fig.~\ref{Fig2res} lead to a clear trace of the omitted terms
in the residuals. Such a reduced form of Eq.~(\ref{AMR}) does not allow
to reproduce the data well, even when a cubic magnetization anisotropy is
included in the SW model (not shown in Fig.~\ref{Fig2res}). The
obtained AMR parameters corresponding to $T = 300$~K (RT)
are shown in Tab.~\ref{Table1}. 

\begin{table}
\begin{tabular}{lll|lll}
AMR Quantity	& 40 nm & 50 nm & 40 nm & 50 nm & AMTP Quantity\\
\hline
%
$\rho_0$ [$\mu \Omega$ cm] & 185.2 & 186.0
&-15.57& -6.61
& $\Sigma_0$ [$\mu $V/K]\\
$a_\mathrm{y2}$ [$\mu \Omega$ cm] & -0.730 & -0.267
&-0.216& -0.248& $s_\mathrm{y2}$ [$\mu $V/K]\\
$a_\mathrm{z2}$ [$\mu \Omega$ cm] & -0.229 & 0.046
&0.270 &0.283& $s_\mathrm{z2}$\ [$\mu $V/K]\\
$a_\mathrm{y4}$ [$\mu \Omega$ cm] & 0.155 & 0.064
&- & - & $s_\mathrm{y4}$\ [$\mu $V/K]
\\
$a_\mathrm{z4}$ [$\mu \Omega$ cm] & -0.242 & -0.106
&- &- & $s_\mathrm{z4}$\ [$\mu $V/K]\\
$a_\mathrm{zy2}$ [$\mu \Omega$ cm] & 0.008 & 0.008
& -& - & $s_\mathrm{zy2}$\ [$\mu $V/K]
\end{tabular}
\caption{Fitted AMR and AMTP parameters at room temperature.}
\label{Table1}
\end{table}

\begin{figure}[h]
\includegraphics[scale=0.6]{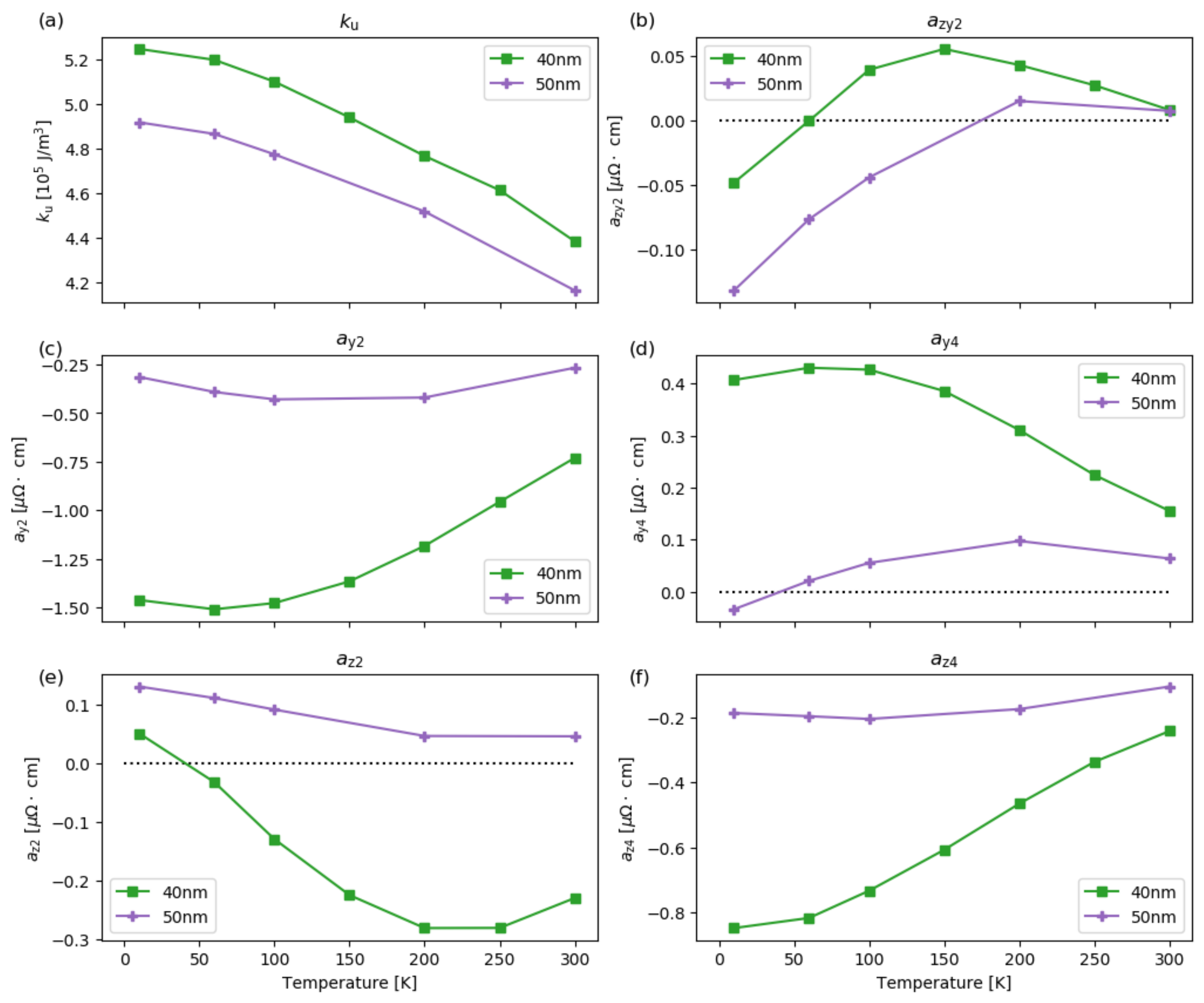}
\caption{Temperature evolution of the parameters obtained by the
  phenomenological fit to AMR data. (A)
  Uniaxial magnetic anisotropy, (B-F) parameters for AMR defined in
  Eq.~\ref{AMR}.}
\label{Fig3}
\end{figure}

Inferred RT values of $k_u$ (438 and 416 kJ/m$^3$ for the 40 and 50 nm
sample, respectively) are in a good agreement with ferromagnetic resonance
measurements carried out independently~\cite{Swekis:2020_a}
and the temperature dependence of such magnetic anisotropy, see
Fig.~\ref{Fig3}(a), is consistent~\cite{Zener:1954_a} with that
of the magnetisation (see Fig.~\ref{Fig1}b). 
Turning our attention to the transport coefficients, the largest of
the AMR parameters is the in-plane lowest-order one: $a_{\mathrm{y2}}$.
It is negative, which reflects that the resistivity is smaller for
$\textbf{m} \parallel \textbf{j}$ than for $\textbf{m} \perp \textbf{j}$,
a situation commonly referred to as {\em negative AMR}~\cite{Tsunoda:2010_a}.
This is opposite to what is found in more common ferromagnets such as iron,
nickel, cobalt and  their alloys~\cite{Turek:2012_a} and more
importantly, it is also consistent with the finding of
Sato~et~al.~\cite{Sato:2019} who found a negative AMR ratio
$(\rho_\parallel -\rho_\perp)/\rho_\perp$ in \ce{Co2MnGa} for current
along \hkl[1 1 0]. Some other ferromagnetic systems, (Ga,Mn)As for
instance~\cite{Rushforth:2007_a}, carry negative AMR too.

Temperature dependences of the AMR parameters are shown in
Fig.~\ref{Fig3}. Their trends for the 50 nm sample are similar to
those of the 40 nm sample except for the out-of-plane lowest-order
$a_\mathrm{z2}$ parameter; even so, the both data sets in Fig.~\ref{Fig3}(e)
seem to have a minimum slightly below RT. According to the absolute
value, the magnitude of the parameters in descending order are as
follows: $|a_{\mathrm{y2}}| > |a_{\mathrm{z4}}| > |a_{\mathrm{z2}}|
>|a_{\mathrm{y4}}| > |a_{\mathrm{zy2}}|$.

Such observations, however, lack any universal validity. 
Trends regarding the
order of magnitude or the sign of the coefficients can be observed,
yet they cannot be generalized. This is also confirmed by results of
similar studies in \ce{Co2FeAl}~\cite{Althammer:2012}  
and in \ce{(Ga, Mn)As}~\cite{Limmer:2006_a,Limmer:2008_a,Althammer:2012}.
Hence, there are always exceptions to a rule: $|a_\mathrm{y2}|$ is usually
the largest of the AMR coefficients (but not for very thin
samples~\cite{Ritzinger:2020_a}) and $a_\mathrm{y4}$ is in most cases positive
(but not at very low temperatures in the 50 nm sample) to give two examples.
Since our model is phenomenological and the microscopic origins of the
AMR mechanisms are not fully understood for \ce{Co2MnGa}, an
explanation of the observed behaviour remains an open question.

Attempts to identify the underlying mechanisms of AMR in related materials have
been undertaken by Kokado and Tsunoda~\cite{unknownREF} whereas the focus
was on electron scattering.
They used a two-current model, taking into account $s$-to-$s$ and
$s$-to-$d$ scattering. The Hamiltonian of the localized d-states includes spin-orbit interaction, an exchange field and a crystal field of cubic or tetragonal symmetry, where the tetragonal distortion is in \hkl[0 0 1] direction. They found that the $a_\mathrm{y4}$ contribution ($C_4$ in their notation) appears under a tetragonal symmetric crystal field, but almost vanishes under cubic symmetry. This is consistent to other studies that reported that a four-fold-contribution ($a_\mathrm{y4}$ in our notation) is not needed to describe the in-plane AMR.
However, thin-films are expected to be strained by the substrate and thus to show some tetragonality, which leads to a non-zero $a_{y4}$ contribution. On the other hand, the strain is different in each sample. Thus, studies that reported a two-fold in-plane AMR (i.e. $a_{y4} = 0$) might have samples with relatively low strain, which are almost cubic.

\subsubsection*{AMTP}

\begin{figure}[h]
\includegraphics[scale=0.6]{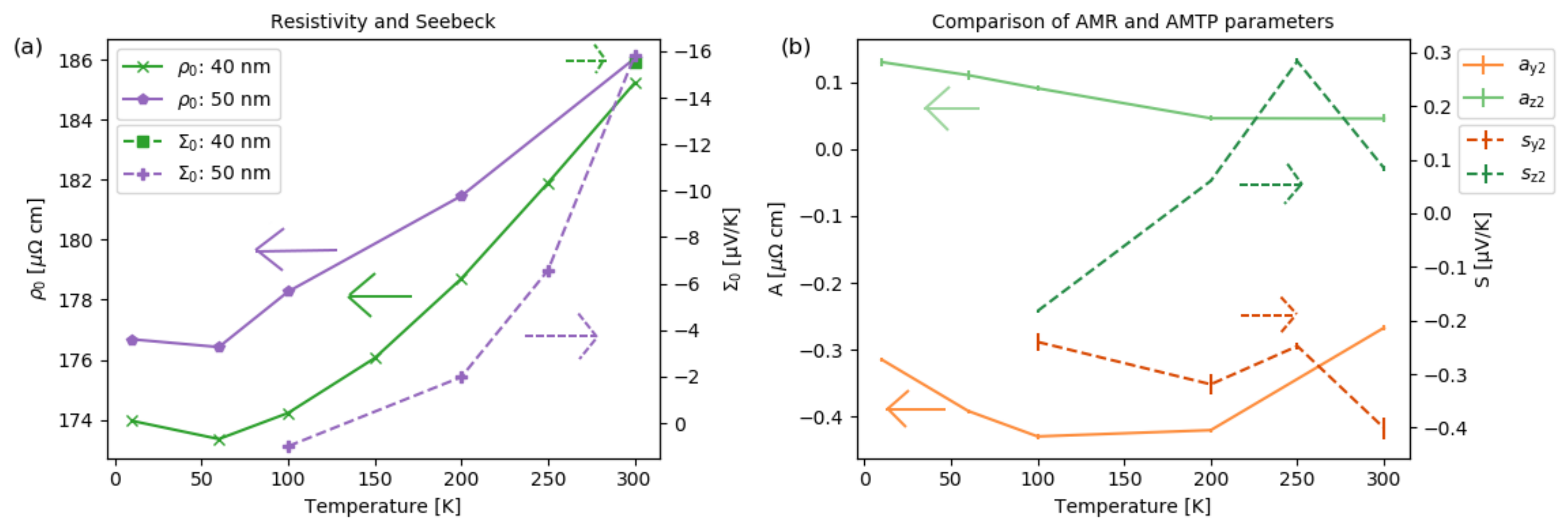}
\caption{a) Longitudinal Seebeck coefficient $\Sigma_0$ as a function
  of the temperature. The different samples are distinguished by the shape
  of symbols. The absolute value of the Seebeck coefficient of the 50
  nm sample is monotonously increasing with temperature. There is a
  sign change at low temperatures. The Seebeck coefficients at $T =
  300$ K for the 40 and 50 nm sample are of comparable magnitude. b)
  Comparison of the AMTP and the corresponding AMR parameters in the
  50 nm sample as a function of temperature. AMR and AMTP are
  distinguished by linestyle. The $m_\mathrm{y}^2$ contributions are
  in shades of orange, the $m_\mathrm{z}^2$ contributions in shades of
  green. They do not seem to follow any common trend. 
  Error bars of AMTP and AMR parameters in b) (implied by the fitting
  procedure) are too small to be visually resolved.}
\label{Fig4}
\end{figure}

To fit the AMTP data, we used a procedure analogous to fitting
AMR except for the anisotropy constant $k_\mathrm{u}$: this parameter has
already been determined before and we now kept it fixed. Note, that
due to the on-chip heating the actual temperature might be slightly
different than indicated. However since the change of $k_\mathrm{u}$
with temperature is small, it does not change the accuracy of our
approach. Given our measurement geometry, $m_xm_ym_z=0$ at all times, hence
Eq.~(\ref{AMTP}) contains the same terms as Eq.~(\ref{AMR}) and in
particular, we started with the lowest-order terms $s_\mathrm{y2}$ and
$s_\mathrm{z2}$. Only these lowest-order parameters and the magnetic
anisotropy were needed to obtain good fits to the AMTP data, which is
a pronounced difference to the AMR.
A reason for this difference could be the noise which is stronger in
the AMTP data as compared to the AMR. We have not been able to achieve
as good resolution as in the case of the AMR. However, the noise
allows us to determine a maximum value of possible higher-order terms,
which need to be smaller than the noise. In absolute terms, the noise is
of the order of magnitude 0.10 $\mu$V/K and below, which implies that
the higher-order symmetries are smaller than about one fifth of the
lowest-order symmetries (see Tab.~\ref{Table1}).
This is not only a striking difference to AMR in our samples, where lower- and higher-order coefficients are in the same order of magnitude, but also to the analysis of AMTP in \ce{(Ga,Mn)As} by Althammer~\cite{Althammer:2012}, where the existence of higher-order AMTP parameter is reported.
In relative terms, the noise shown in Fig.~\ref{Fig2res} (as residuals
after subtracting the fits from experimental data) is large which is a
consequence of difficulties in controlling the temperature gradient
under experimental conditions. The temperature evolution of the AMTP
parameters as well as a comparison to the lowest-order AMR parameters
is shown in Fig.~\ref{Fig4}(b).

In Fig. \ref{Fig4}(a), the Seebeck coefficient $\Sigma_0$ is shown as
function of temperature. 
In literature, the Seebeck coefficient of \ce{Co2MnGa} ranges
between approximately $-2 \mu$V/K and $-30 \mu$V/K at RT, whereas no clear
trend is recognizable and the present measurements fall within this range:
we find $\Sigma_0$ for both samples close to $-15 \mu$V/K. Seebeck
coefficient of the 50~nm sample is increasing in absolute value with
increasing temperature, as expected from previously published
experiments~\cite{Guin2019,Balke:2010_a}.

 As mentioned in Sec. I, the sign of $\Sigma_0$ is attributed to the dominant charge carrier type, which are electrons for $\Sigma_0 < 0$. A negative Seebeck coefficient is reported not only in \ce{Co2MnGa}, but in Co-based Heusler compounds in general~\cite{Hu:2020_a}. The sign change of $\Sigma_0$ would mean that the dominant charge carrier type switches from electrons to holes for decreasing temperatures. This was shown to occur in other materials depending on the doping~\cite{Lue:2002_a}.
However, other scenarios are also plausible which we discuss in the next
subsection and we now turn our attention to the AMTP.

We notice that neither $s_\mathrm{y2}$ nor $s_\mathrm{z2}$, whose
temperature dependence is shown in Fig.~\ref{Fig4}(b), can be correlated
to their AMR counterparts.
While $a_\mathrm{z2}$ is decreasing with increasing temperature and the dependence of $a_\mathrm{y2}$ is non-monotonic, the AMTP parameters exhibit very different dependence, although they were measured on exactly same device in the and same structure. 
At $T = 250$ K in both AMTP parameters, there is an outlier. This
might be due to a thermal offset variation. Together with the sign
change of $s_\mathrm{z2}$ the lack of correlation between the AMR and
AMTP is evident. This is in striking contrast to the transversal
transport measured in the same material \cite{Park2020}. In that study
the AHE and ANE were measured simultaneously and clear correlation was
observed including the presence of outliers.

From our study it appears that comparing the lowest-order parameters,
no common trend can be found between the AMR parameters and their
respective AMTP counterpart when looking at the temperature
dependence. Moreover, in case of the ZX rotation (see
Fig.~\ref{Fig2}f), even the raw angular sweep data are clearly very different:
the fourth-order terms can be basically seen by naked eye in the AMR while in
the AMTP, they are apparently absent and no such terms can
be the identified in the residuals on right panels of Fig.~\ref{Fig2res}.
In the systematic study of AMR and AMTP in \ce{(Ga, Mn) As} by
Althammer, higher-order contributions have been found for both effects, but
the parameters did not appear to be correlated with each other. Since
the amount of data in this and all past studies is relatively small,
further studies are desirable to investigate if any correlation
between AMR und AMTP coefficients exists.
%

The fact, that in two different systems the AMR and the AMTP follow
different trends is, however, a strong indication, that there might be
a more fundamental reason behind this discrepancy such as the
suppressed role of anisotropic scattering in the AMTP, which calls for
further investigation; \myhighlight{below, we discuss one possible direction of
such tentative research.}

\subsubsection*{Mott rule and phonon drag}

Electron and heat transport linear response coefficients are tensorial
quantities which obey Onsager relations~\cite{Smrcka:1976_a} and
combine into the Seebeck coefficient $\Sigma_0$. To see this, we
rewrite the electron transport equation as
$$
  \textbf{E}=\rho\textbf{j}+\Sigma_0\nabla T
$$
and set $\textbf{j}=0$. Under the assumption of elastic scattering, the
Mott rule holds (still in tensorial sense)
\begin{equation}
  \Sigma_0=-eL_0T\rho\sigma'=eL_0T\rho'\sigma
  \label{MOTT}
\end{equation}
where $\rho\sigma=1$ and the prime denotes the derivative with respect
to energy $E$ at chemical potential. For semiconductors, this derivative
can be straightforwardly accessed through carrier-density-dependent
conductivity~\cite{Miyasato:2007_a} but no such possibility is obvious
in our metallic system. However, when resistivity is broken down into
a magnetisation-independent background $\rho_0$ and a small correction
$\Delta\rho$ that effectively constitutes the AMR, the following
observation is possible.

Let us first assume that $\rho$ which depends both on $E$ and $\textbf{M}$
can be written as
\begin{equation}\label{Eq-ass}
  \rho(E,\textbf{M})=\rho_0(E)+\Delta\rho(\textbf{M}).
\end{equation}
Eq.~(\ref{MOTT}) can then be rewritten as $\Sigma_0=eL_0T\rho_0'\sigma$
and since $\rho_0$ is a scalar quantity, it essentially implies that
AMTP is proportional to the AMR (tensors $\Sigma_0$ and $\sigma$ are
proportional through a scalar factor). In terms of data shown in
Fig.~\ref{Fig2},
this could seem to imply that dominant non-crystalline contributions
of AMR (XY and ZY rotations) indeed follow while the more delicate and
smaller crystalline terms (ZX rotation) break the assumption~(\ref{Eq-ass})
and give different AMR and AMTP angular dependences.

On a careful inspection however, we notice that AMR and AMTP
amplitudes in the left and middle panels of Fig.~\ref{Fig2} are not
proportional the same way as the average $\rho$ and $\Sigma_0$ (this
is made explicit by scaling of the vertical axes on the left three
panels in Fig.~\ref{Fig2res}). While
the assumption~(\ref{Eq-ass}) may be not accurately fulfilled (specifically,
the $\Delta\rho$ term is unlikely to be perfectly energy-independent),
there is also another possibility to explain this discrepancy. We note
that thermopower is strongly temperature-dependent (data in Fig.~\ref{Fig4}a
show that it changes almost by a factor of three between 250 and
300~K) and at the same time, $\rho_0$ changes only at the order of per cent
in the same temperature interval.


This explanation is related to the phonon drag contribution to
thermopower which is not included in the Mott formula~(\ref{MOTT}). In
semiconductors, this contribution can easily exceed (see Fig.~12 in
Protik\&Broido~\cite{Protik:2020_a}) the electronic thermopower or, in
other words, it can reach the level of 100~$\mu$V/K. It is therefore
plausible that the measured (relatively small) value of $\Sigma_0$ is
the result of competition of two (large, relative to $\Sigma_0$)
contributions: the usual electronic contribution related to the
Fermi-Dirac factor $f_{FD}(E)$ depending on temperature and the phonon
drag contribution caused by electron-phonon interaction.
Once the phonon drag contribution would be removed from
the measured $\Sigma_0$, the amplitude-to-average ratio of AMTP drops
to the same level as for the AMR (here, we again refer to the
Fig.~\ref{Fig2}(b and (d) consistent with proportionality of
AMR and AMTP implied by the Mott formula under assumption~(\ref{Eq-ass}).
This way, our measurements suggest a sizable phonon drag effect in the
thermopower of \ce{Co2MnGa}. Alternatively, magnon drag could be at
works~\cite{Watzman:2016_a,Polash:2020_a}. Regarding
Fig.~\ref{Fig2}(f) we note that quite clearly, proportionality between
AMR and AMTP is by no means exact and it seems (in the view of data in
Fig.~\ref{Fig4}(b) that thermopower in \ce{Co2MnGa} is more sensitive
to crystallographic orientation than resistivity.

\subsection*{V. SUMMARY}

In this study, we compared the AMR and AMTP in two
\ce{Co2MnGa}-thin-film samples using a simple Free Energy density and
phenomenological symmetry-based models for AMR and AMTP based on a
series expansion in powers of the magnetization direction vector
$\textbf{m}$. We showed that non-zero resistivity-contributions up to
4th order are necessary for a sensible modeling, where in the AMTP
only lowest-order contributions are necessary. The AMR and AMTP are
not showing any trends in common, which is consistent with previous
studies in \ce{(Ga, Mn)As}. We experimentally confirm presence of a
crystalline contribution to the AMTP.
It appears that the universal Mott rule validity is broken due to a
discrepancy of the symmetries of AMR and AMTP in one rotation plane.
This discrepancy was discussed in terms of a significant phonon (magnon) drag
contribution to thermopower, which might be the origin of such a
discrepancy.


The results of this study call for further enquiry: First of all, we need to broaden our understanding about the origins and governing influences in AMR, but also in AMTP. Theoretical studies discussing influences in AMTP similar to whose about AMR are desirable. Experimental studies using sets of samples which are systematic with respect to strain, composition or other influences can help us also along this way. \\

\section*{Acknowledgements}

We thank Peter Swekis for discussions and Juliane Scheiter
for technical support with magnetometer. HR acknowledges support from 
Christiane N\"usslein-Volhard Stiftung and funding through
W\"urzburg--Dresden Cluster of Excellence, SFB 1143 (project ID 247310070)
and FET Open RIA Grant No. 766566 also deserves a grateful mention.

\begin{appendix}

\section{Notes on derivation of Eqs.~\ref{AMR},\ref{AMTP}}

Comments on the full form of $\rho^{(i)}$ in Eq.~(\ref{Tensor}) are
given here, from which Eq.~(\ref{AMR}) can be derived. A more detailed
description can be found in Appendix A.1 of thesis~\cite{Ritzinger:2020_a}
of one of the authors. Only terms corresponding to cubic symmetry are included
and summation over repeated indexes (Einstein notation) is implied. Note that the underlying coordinate system is equivalent to the one applied in Eq. \ref{Tensor}, thus: $m_{\mathrm{n}} \in \{m_{\hkl[1 0 0]}, m_{\hkl[0 1 0]}, m_{\hkl[0 0 1]} \}$.
We skip $\rho_{\mathrm{ijk}}^{(1)}$ which does not appear in
$\rho_\mathrm{long}$ the in-plane geometry (it corresponds to the
anomalous Hall effect). The lowest non-trivial order thus becomes
$$
\rho_{\mathrm{ijkl}}^{(2)} m_{\mathrm{k}}m_{\mathrm{l}}=
C_1\left(\begin{array}{ccc} m_{\mathrm{x}}^2 & 0 & 0 \\
  0 & m_{\mathrm{y}}^2 & 0 \\
  0 & 0 & m_{\mathrm{z}}^2
\end{array}\right)+
C_2\left(\begin{array}{ccc} 0 & m_{\mathrm{x}}m_{\mathrm{y}}
  & m_{\mathrm{x}}m_{\mathrm{z}} \\
  m_{\mathrm{x}}m_{\mathrm{y}} & 0 & m_{\mathrm{y}}m_{\mathrm{z}} \\
  m_{\mathrm{x}}m_{\mathrm{z}} & m_{\mathrm{y}}m_{\mathrm{z}} & 0
\end{array}\right).
$$
The third-order terms, $\rho_{\mathrm{ijklm}}^{(3)}$, again do not
contribute in Eq.~(\ref{Proj}), and the fourth-order terms
$\rho_{\mathrm{ijklmn}}^{(4)} m_{\mathrm{k}}m_{\mathrm{l}}
m_{\mathrm{m}}m_{\mathrm{n}}$ become
$$
E_1\left(\begin{array}{ccc} m_{\mathrm{x}}^4 & 0 & 0 \\
  0 & m_{\mathrm{y}}^4 & 0 \\
  0 & 0 & m_{\mathrm{z}}^4
\end{array}\right)+
E_2\left(\begin{array}{ccc} m_{\mathrm{y}}^2m_{\mathrm{z}}^2 & 0 & 0 \\
  0 & m_{\mathrm{x}}^2m_{\mathrm{z}}^2 & 0 \\
  0 & 0 & m_{\mathrm{x}}^2m_{\mathrm{y}}^2
\end{array}\right)+
E_3\left(\begin{array}{ccc} 0 & m_{\mathrm{x}}m_{\mathrm{y}}m_{\mathrm{z}}^2
  & m_{\mathrm{x}}m_{\mathrm{z}}m_{\mathrm{y}}^2 \\
  m_{\mathrm{x}}m_{\mathrm{y}}m_{\mathrm{z}}^2 & 0 &
  m_{\mathrm{y}}m_{\mathrm{z}}m_{\mathrm{x}}^2 \\
  m_{\mathrm{x}}m_{\mathrm{z}}m_{\mathrm{y}}^2 &
  m_{\mathrm{y}}m_{\mathrm{z}}m_{\mathrm{x}}^2 &0
\end{array}\right).
$$
%
%
%
Additional terms appear when lower symmetry is assumed, which is in our case due to the tetragonal distortion along the $\hat{z} = \hkl[0 0 1]$ axis of the thin-film samples. The following zeroth-order and second-order terms add to the resistivity tensor in the case of tetragonal symmetry:
$$
\left(\begin{array}{rrr}
0 & 0 & 0 \\
0 & 0 & 0 \\
0 & 0 & a \\
\end{array} \right)
+\left(\begin{array}{ccc}
c_3 m_\mathrm{z}^2 & 0 & c_2 m_\mathrm{x}m_\mathrm{z} \\
0 & c_3 m_\mathrm{z}^2 & c_2 m_\mathrm{y}m_\mathrm{z} \\
c_2 m_\mathrm{x}m_\mathrm{z} & c_2 m_\mathrm{y}m_\mathrm{z} & c_1 m_\mathrm{z}^2 \\
\end{array} \right)
$$
Under tetragonal symmetry, the fourth-order terms are supplemented by:
$$\left(\begin{array}{ccc}
e_2 m_\mathrm{y}^2 m_\mathrm{z}^2 + e_4 m_\mathrm{z}^4 & e_3 m_\mathrm{x} m_\mathrm{y} m_\mathrm{z}^2 &
e_6 m_\mathrm{x} m_\mathrm{y}^2 m_\mathrm{z} + e_7 m_\mathrm{x} m_\mathrm{z}^3 \\
e_3 m_\mathrm{x} m_\mathrm{y} m_\mathrm{z}^2 & e_2 m_\mathrm{x}^2 m_\mathrm{z}^2 + e_4 m_\mathrm{z}^4 & e_6 m_\mathrm{x}^2 m_\mathrm{y} m_\mathrm{z} + e_7 m_\mathrm{y} m_\mathrm{z}^3  \\
e_6 m_\mathrm{x} m_\mathrm{y}^2 m_\mathrm{z} + e_7 m_\mathrm{x} m_\mathrm{z}^3 &
e_6 m_\mathrm{x}^2 m_\mathrm{y} m_\mathrm{z} + e_7 m_\mathrm{y} m_\mathrm{z}^3 & e_5 m_\mathrm{x}^2 m_\mathrm{y}^2 + e_1 m_\mathrm{z}^4 \\
\end{array} \right)
$$


In a similar (yet distinct) manner, the Seebeck tensor can be expanded,
see Appendix B.1 of Ref.~\cite{Ritzinger:2020_a}. In the case of tetragonal
symmetry, to which Eq.~(\ref{AMTP}) applies, several additional terms
appear but they do not contribute to $\Sigma_\mathrm{yy}$ except for
$$
D_7m_{\mathrm{x}}m_{\mathrm{y}}m_{\mathrm{z}} \left(\begin{array}{ccc}
  -1& 0 & 0 \\
  0 & 1 & 0 \\
  0 & 0 & 0
  \end{array}\right)
$$
which gives rise to the last term in Eq.~(\ref{AMTP}). Therefore even in
our setup, $\rho_\mathrm{yy}$ and $\Sigma_\mathrm{yy}$ allow in principle
for a different functional form albeit not with our constraint to XY,
YZ, and XZ rotations of magnetic field.
Further information can be found in Ref.~\cite{Althammer:2012} or in
the Appendix of Ref.~\cite{Ritzinger:2020_a}.

\end{appendix}


\begin{thebibliography}{99}%
\makeatletter
\providecommand \@ifxundefined [1]{%
 \@ifx{#1\undefined}
}%
\providecommand \@ifnum [1]{%
 \ifnum #1\expandafter \@firstoftwo
 \else \expandafter \@secondoftwo
 \fi
}%
\providecommand \@ifx [1]{%
 \ifx #1\expandafter \@firstoftwo
 \else \expandafter \@secondoftwo
 \fi
}%
\providecommand \natexlab [1]{#1}%
\providecommand \enquote  [1]{``#1''}%
\providecommand \bibnamefont  [1]{#1}%
\providecommand \bibfnamefont [1]{#1}%
\providecommand \citenamefont [1]{#1}%
\providecommand \href@noop [0]{\@secondoftwo}%
\providecommand \href [0]{\begingroup \@sanitize@url \@href}%
\providecommand \@href[1]{\@@startlink{#1}\@@href}%
\providecommand \@@href[1]{\endgroup#1\@@endlink}%
\providecommand \@sanitize@url [0]{\catcode `\\12\catcode `\$12\catcode
  `\&12\catcode `\#12\catcode `\^12\catcode `\_12\catcode `\%12\relax}%
\providecommand \@@startlink[1]{}%
\providecommand \@@endlink[0]{}%
\providecommand \url  [0]{\begingroup\@sanitize@url \@url }%
\providecommand \@url [1]{\endgroup\@href {#1}{\urlprefix }}%
\providecommand \urlprefix  [0]{URL }%
\providecommand \Eprint [0]{\href }%
\providecommand \doibase [0]{http://dx.doi.org/}%
\providecommand \selectlanguage [0]{\@gobble}%
\providecommand \bibinfo  [0]{\@secondoftwo}%
\providecommand \bibfield  [0]{\@secondoftwo}%
\providecommand \translation [1]{[#1]}%
\providecommand \BibitemOpen [0]{}%
\providecommand \bibitemStop [0]{}%
\providecommand \bibitemNoStop [0]{.\EOS\space}%
\providecommand \EOS [0]{\spacefactor3000\relax}%
\providecommand \BibitemShut  [1]{\csname bibitem#1\endcsname}%
\let\auto@bib@innerbib\@empty

\bibitem{Thomson:1857_a} W. Thomson,
  Proc. R. Soc. Lond. \textbf{8}, 546-550 (1856)

\bibitem{Wang:2020_a} M. Wang, C. Andrews, S. Reimers, O. J. Amin, P. Wadley, R. P. Campion, S. F. Poole, J. Felton, K. W. Edmonds, B. L. Gallagher, A. W. Rushforth, O. Makarovsky, K. Gas, M. Sawicki, D. Kriegner, J. Zubáč, K. Olejník, V. Novák, T. Jungwirth, M. Shahrokhvand, U. Zeitler, S. S. Dhesi, and F. Maccherozzi, Phys. Rev. B \textbf{101}, 094429 (2020)


\bibitem{Volny:2020_a} J. Volný, D. Wagenknecht, J. Železný, P. Harcuba, E. Duverger–Nedellec, R. H. Colman, J. Kudrnovský, I. Turek, K. Uhlířová, and K. Výborný, Phys. Rev. Materials \textbf{4}, 064403 (2020). 

\bibitem{Miao:2020_a} Y. Miao, X. Chen and D.--S. Xue,
  J. Magn. Magn. Mat. 512, 167013 (2020).

  
\bibitem{Wesenberg:2018_a} D. Wesenberg, A. Hojem, R. K. Bennet and B. L. Zink, J. Phys. D: Appl. Phys. \textbf{51} 244005 (2018)

\bibitem{Yu:2017_a} H. Yu, S. D. Brechet, J.-P. Ansermet, Phys. Lett. A \textbf{381}, 825-837 (2017)


\bibitem{Hu:2020_a} Fig. 8 and Tab. 2 in J. Hu, S. Granville, H. Yu, Ann. Phys., Lpz. \textbf{532}, 1900456 (2020)


\bibitem{Sakai2018} A. Sakai, Y. P. Mizuta, A. A. Nugroho, R. Sihombing, T. Koretsune, M.-T. Suzuki, N. Takemori, R. Ishii, D. Nishio-Hamane, R. Arita, P. Goswami and S. Nakatsuji, Nat. Phys. \textbf{14}, 1119–1124 (2018)


\bibitem{Reichlova2018} H. Reichlova, R. Schlitz, S. Beckert, P. Swekis, A. Markou, Y.-C. Chen, D. Kriegner, S. Fabretti, G. H. Park, A. Niemann, S. Sudheendra, A. Thomas, K. Nielsch, C. Felser, and S. T. B. Goennenwein, Appl. Phys. Lett. \textbf{113}, 212405 (2018)



\bibitem{Belopolski2019} I. Belopolski, K. Manna, D. S. Sanchez, G. Chang, B. Ernst, J. Yin, S. S. Zhang, T. Cochran, N. Shumiya, H. Zheng, B. Singh, G. Bian, D. Multer, M. Litskevich, X. Zhou, S.-M. Huang, B. Wang, T.-R. Chang, S.-Y. Xu, A. Bansil, C. Felser, H. Lin, and M. Z. Hasan, Science \textbf{365}, 1278 (2019).

\bibitem{Park2020} G.-H. Park, H. Reichlova, R. Schlitz, M. Lammel, A. Markou, P. Swekis, P. Ritzinger, D. Kriegner, J. Noky, J. Gayles, Y. Sun, C. Felser, K. Nielsch, S. T. B. Goennenwein, and A. Thomas, Phys. Rev. B \textbf{101} (2020)

\bibitem{Graf:2011_a} T. Graf, C. Felser, S. S. P. Parkin, Prog. Solid. State Ch. \textbf{39}, 1-50 (2011)


\bibitem{Protik:2020_a} N. H. Protik and D. A. Broido, Phys. Rev. B \textbf{101}, 075202 (2020) 

\bibitem{Polash:2020_a} Md. M. H. Polash, F. Mohaddes, M. Rasoulianboroujeni, D. Vashaee, J. Mater. Chem. C \textbf{8}, 4049-4057 (2020)

\bibitem{Rushforth:2007_a} A. W. Rushforth, K. Výborný, C. S. King, K. W. Edmonds, R. P. Campion, C. T. Foxon, J. Wunderlich, A. C. Irvine, P. Vašek, V. Novák, K. Olejník, Jairo Sinova, T. Jungwirth, and B. L. Gallagher, Phys. Rev. Lett. \textbf{99}, 147207 (2007)  

\bibitem{Doring:1938_a} W. D\"oring, Ann. Phys., Lpz. \textbf{424}, 259 (1938) 

\bibitem{Ranieri:2008_a} E. De Ranieri, A. W. Rushforth, K. Výborný, U. Rana, E. Ahmad, R. P. Campion, C. T. Foxon, B. L. Gallagher, A. C. Irvine, J. Wunderlich, New J. Phys. \textbf{10}, 065003 (2008)

\bibitem{Ky:1966_a} V. D. Ky, Phys. Status Solidi B \textbf{17}, K207 (1966).

\bibitem{Snyder:2008} G. J. Snyder and E. S. Toberer, Nat. Mater. \textbf{7}, 105–114 (2008) 

\bibitem{Janda:2020_b} T. Janda, J. Godinho, T. Ostatnicky,
  E. Pfitzner, G. Ulrich, A. Hoehl, S. Reimers, Z. Šobá\v n, T. Metzger,
H. Reichlová, V. Novák, R. P. Campion, J. Heberle, P. Wadley, K. W. Edmonds, O. J. Amin, J. S. Chauhan, S. S. Dhesi,
F. Maccherozzi, R. M. Otxoa, P. E. Roy, K. Olejník, P. Nemec, T. Jungwirth, B. Kaestner, and J. Wunderlich, Phys. Rev. Materials \textbf{4}, 094413 (2020)

\bibitem{Loevvik:2020_a} O.M.L{\o}vvik, Espen Flage-Larsen, and Gunstein Skomedal, J. Appl. Phys. \textbf{128}, 125105 (2020)

\bibitem{note2} \myhighlight{Electronic bands in the bulk can be characterised
by topological invariants which indicate whether or not there will be surface
states. These are related to Berry curvature, a quantity related to
ANE~\cite{Sakai2018},  through 
Eq.~2 of J. K\"ubler and C. Felser, Europhys. Lett. \textbf{114}, 47005 (2016).}



\bibitem{Sakai2020} A. Sakai, S. Minami, T. Koretsune, T. Chen, T. Higo, Y. Wang, T. Nomoto, M. Hirayama, S. Miwa, D. Nishio-Hamane, F. Ishii, R. Arita and S. Nakatsuji, Nature \textbf{581}, 53 (2020)


\bibitem{Limmer:2008_a} W. Limmer, J. Daeubler, L. Dreher, M. Glunk, W. Schoch, S. Schwaiger, and R. Sauer,
Phys. Rev. B \textbf{77}, 205210 (2008). 

\bibitem{Althammer:2012} M. Althammer, doctoral thesis, Technische Universität München, 2012  

\bibitem{Reimer:2017_a} O. Reimer, D. Meier, M. Bovender, L. Helmich, J.-O. Dreessen, J. Krieft, A. S. Shestakov, C. H. Back, J.-M. Schmalhorst, A. Hütten, G. Reiss and T. Kuschel, Sci. Rep. \textbf{7}, 40586 (2017)


\bibitem{Jayathilaka:2015_a} P.B.Jayathilaka, D.D.Belyea, T.J.Fawcett, 
  and Casey W.Miller,
  J. Magn. Magn. Mat. 382, 376 (2015).
  
\bibitem{Janda:2020_a} T. Janda, J. Godinho, T. Ostatnicky, E. Pfitzner, G. Ulrich, A. Hoehl, S. Reimers, Z. Šobán, T. Metzger, H. Reichlová, V. Novák, R. P. Campion, J. Heberle, P. Wadley, K. W. Edmonds, O. J. Amin, J. S. Chauhan, S. S. Dhesi,
F. Maccherozzi, R. M. Otxoa, P. E. Roy, K. Olejník, P. Nemec, T. Jungwirth, B. Kaestner, and J. Wunderlich, Phys. Rev. Materials \textbf{4}, 094413 (2020)


\bibitem{Avery:2012_a} A. D. Avery, M. R. Pufall, and B. L. Zink, Phys. Rev. Lett. \textbf{109}, 196602 (2012) 

\bibitem{Markou:2019} A. Markou, D. Kriegner, J. Gayles, L. Zhang, Y.-C. Chen, B. Ernst, Y.-H. Lai, W. Schnelle, Y.-H. Chu, Y. Sun, and C. Felser, Phys. Rev. B \textbf{100}, 054422 (2019)  

\bibitem{Pietsch:2004} U. Pietsch, V. Hol\'y, and T. Baumbach, High-resolution
X-ray Scattering: From Thin Films to Lateral Nanostructures
(Springer, New York, 2004). 

\bibitem{Limmer:2006_a} W. Limmer, M. Glunk, J. Daeubler, T. Hummel, W. Schoch, R. Sauer, C. Bihler, H. Huebl, M. S. Brandt, and S. T. B. Goennenwein, Phys. Rev. B \textbf{74}, 205205 (2006) 

\bibitem{SW48} E. C. Stoner and E. P. Wohlfarth, Phil. Trans. Roy. Soc. A \textbf{240}, 599 (1948).

\bibitem{Swekis:2020_a} 
P. Swekis, A. S. Sukhanov, Y.-C. Chen, A. Gloskovskii,
G. H. Fecher, I. Panagiotopoulos, V. Ukleev, A.
Devishvili, A. Vorobiev, D. S. Inosov, S.
T. B. Goennenwein, C. Felser and A. Markou
(unpublished) 


\bibitem{note3} A tensor representing a macroscopic physical property
of a crystal must be invariant under all symmetry operations of the
corresponding point-group.


\bibitem{Zener:1954_a} C. Zener,
Phys. Rev. \textbf{96}, 1335 (1954). 

\bibitem{Tsunoda:2010_a} M. Tsunoda, H. Takahashi, S. Kokado, Y. Komasaki, A. Sakuma and M. Takahashi, Appl. Phys. Express \textbf{3}, 113003 (2010)

\bibitem{Turek:2012_a} I. Turek, J. Kudrnovský, and V. Drchal,
Phys. Rev. B \textbf{86}, 014405 (2012). 

\bibitem{Sato:2019} T. Sato, S. Kokado, M. Tsujikawa, T. Ogawa, S. Kosaka, M. Shirai and M. Tsunoda, Appl. Phys. Express \textbf{12}, 103005 (2019)    


\bibitem{Ritzinger:2020_a} P. Ritzinger, master thesis, Technische Universität Dresden, 2020 

\bibitem{unknownREF} S. Kokado and M. Tsunoda,
J. Phys. Soc. Jpn. \textbf{84}, 094710 (2015). 

\bibitem{Guin2019} S. N. Guin, K. Manna, J. Noky, S. J. Watzman, C. Fu, N. Kumar, W. Schnelle, C. Shekhar, Y. Sun, J. Gooth, and C. Felser, NPG Asia Mater. \textbf{11}:16 (2019)

\bibitem{Balke:2010_a} B. Balke, S. Ouardia, T. Graf, J. Barth, C. G. F. Blum, G. H. Fecher, A. Shkabko, A. Weidenkaff, C. Felser,
Sol. St. Comm. \textbf{150}, 529 (2010).

\bibitem{Lue:2002_a} C. S. Lue and Y.-K. Kuo,
Phys. Rev. B \textbf{66}, 085121 (2002).   

\bibitem{Smrcka:1976_a} L. Smrčka and P. Vašek, Czech. J. Phys. B \textbf{26}, 1137--1147 (1976)

\bibitem{Miyasato:2007_a} T. Miyasato, N. Abe, T. Fujii, A. Asamitsu, S. Onoda, Y. Onose, N. Nagaosa, and Y. Tokura, Phys. Rev. Lett. \textbf{99}, 086602 (2007) 


\bibitem{Watzman:2016_a} S. J. Watzman, R. A. Duine, Y. Tserkovnyak, S. R. Boona, H. Jin, A. Prakash, Y. Zheng, and J. P. Heremans,
Phys. Rev. B \textbf{94}, 144407 (2016) 


\end{thebibliography}
\end{document}